\documentclass{article}

\PassOptionsToPackage{numbers}{natbib}
\usepackage[final]{neurips_2024}

\usepackage[utf8]{inputenc} 
\usepackage[T1]{fontenc}    
\usepackage{hyperref}       
\usepackage{url}            
\usepackage[table]{xcolor}
\usepackage{booktabs}       
\usepackage{amsfonts}       
\usepackage{nicefrac}       
\usepackage{microtype}      
\usepackage{xcolor}         
\usepackage{tikz}
\usepackage{amsmath} 
\usepackage{hyperref}
\usepackage{wrapfig}
\usepackage{siunitx}
\usepackage{multirow}
\usepackage{graphicx}
\usepackage{threeparttable} 
\usepackage{caption}
\usepackage[subtle]{savetrees}
\definecolor{lightgray}{gray}{0.9}
\definecolor{soma}{RGB}{16, 38, 124}
\definecolor{def}{RGB}{67, 150, 42}
\definecolor{vis}{RGB}{238,142,79}
\newcommand*\circledwhite[1]{\tikz[baseline=(char.base)]{\node[shape=circle,draw,inner sep=0.25pt,fill=black,text=white] (char) {#1};}}

\title{Brain-JEPA: Brain Dynamics Foundation Model with Gradient Positioning and Spatiotemporal Masking}

\author{
  \textbf{Zijian~Dong\thanks{\emph{Equal contribution}}, \ \ Ruilin~Li\footnotemark[1], \ \ Yilei Wu, \ \ Thuan Tinh Nguyen, \ \ Joanna Su Xian Chong, \ \ Fang Ji,} \\[5pt]
  \textbf{Nathanael Ren Jie Tong, \ \ Christopher Li Hsian Chen, \ \ Juan Helen Zhou\thanks{\emph{Corresponding author}}} \\[8pt]
  \texttt{zijian.dong@u.nus.edu, \{li.rl, helen.zhou\}@nus.edu.sg}
  \\[8pt]
  National University of Singapore
}

\begin{document}

\maketitle

\begin{abstract}
  We introduce \emph{Brain-JEPA}, a brain dynamics foundation model with the Joint-Embedding Predictive Architecture (JEPA). This pioneering model achieves state-of-the-art performance in demographic prediction, disease diagnosis/prognosis, and trait prediction through fine-tuning. Furthermore, it excels in off-the-shelf evaluations (\emph{e.g.}, linear probing) and demonstrates superior generalizability across different ethnic groups, surpassing the previous large model for brain activity significantly. Brain-JEPA incorporates two innovative techniques: \textbf{Brain Gradient Positioning} and \textbf{Spatiotemporal Masking}. Brain Gradient Positioning introduces a functional coordinate system for brain functional parcellation, enhancing the positional encoding of different Regions of Interest (ROIs). Spatiotemporal Masking, tailored to the unique characteristics of fMRI data, addresses the challenge of heterogeneous time-series patches. These methodologies enhance model performance and advance our understanding of the neural circuits underlying cognition. Overall, Brain-JEPA is paving the way to address pivotal questions of building brain functional coordinate system and masking brain activity at the AI-neuroscience interface, and setting a potentially new paradigm in brain activity analysis through downstream adaptation. \textit{Code is available at: \href{https://github.com/Eric-LRL/Brain-JEPA}{https://github.com/Eric-LRL/Brain-JEPA.}}
\end{abstract}

\section{Introduction}
\label{intro}

Understanding large-scale brain activity data is crucial for deciphering the complex mechanisms underlying cognitive processes and human behavior. Functional magnetic resonance imaging (fMRI) captures blood-oxygen-level dependent (BOLD) signals that reflect regional brain activity. It emerges as an indispensable tool in neuroscience for identifying the neural bases of cognitive processes \citep{logothetis2008we,heeger2002does,logothetis2001neurophysiological}. Deep learning approaches have been developed for fMRI analysis, improving brain disease diagnosis and deepening insights into cognition and behavior \citep{kan2022brain,li2021braingnn,kawahara2017brainnetcnn,cui2022braingb,drysdale2017resting,iidaka2015resting,dong2024prompt,dong2023beyond,wu2023mixup,10021043}. Despite notable advances, these task-specific models suffer from limited generalizability and adaptability to other tasks. In addition, they fail to leverage the vast amounts of unlabeled fMRI data available \citep{caro2023brainlm,wen2023graph}.

Artificial intelligence (AI) is experiencing a paradigm shift from task-specific training to building foundation models that are trained on extensive data using self-supervision at scale \citep{bommasani2021opportunities}. Unlike the models with singular functions, foundation models can be adapted to a diverse array of downstream tasks. Large language models such as GPT \citep{achiam2023gpt} and LLaMA \citep{touvron2023llama} have shown significant potential in natural language processing, with expansive applications in healthcare, biomedicine, and beyond \citep{bommasani2021opportunities}. 

In the field of fMRI time series analysis, brain language model (brainLM) is one of the most representative foundation models \citep{caro2023brainlm}. It is a masked autoencoder (MAE) \citep{He2022MAE} trained to reconstruct masked fMRI time series. However, as an indirect measure of neuronal activity, the BOLD signal has a relatively low signal-to-noise ratio (SNR), which is influenced by a mixture of factors and distorted by non-neuronal fluctuations \citep{caballero2017methods}. Filling every bit of the fMRI time series as in brainLM can hinder the model's ability to distinguish between noise and actual signals. This can result in either amplifying noise or missing critical subtle variations in brain activity. Unlike natural images, which have high information density with structures such as edges and colors, fMRI data has spatiotemporally sparser signals distributed across brain volumes without clear boundaries, making it difficult to accurately reconstruct signal of masked regions of interest (ROIs). Furthermore, it has been widely shown that masked pretraining in generative architectures such as MAE leads to suboptimal performance in off-the-shelf evaluations (\textit{e.g.}, linear probing) \citep{assran2023self}. Because of that, BrainLM requires computationally intensive end-to-end finetuning, with a three-layer MLP attached to the pretrained encoder, to achieve optimal performance. Furthermore, the absence of comparisons with state-of-the-art methods for downstream task performance and the focus only on Caucasian cohorts limit BrainLM's applicability in clinical settings. 

Therefore, rather than focusing on the original brain activity time series, the inherent noise and sparse information density of fMRI lead us to explore the latent space of fMRI time series extracted from a strong encoder (\emph{e.g.}, Vision Transformer (ViT) \citep{dosovitskiy2020image}). It potentially offers a higher SNR after "compression", achieving a greater level of abstraction that captures subtle yet crucial patterns \citep{lecun2022path}. Recently, Imaged-based Joint-Embedding Predictive Architecture (I-JEPA) has been proposed as a non-generative architecture for self-supervised learning from images \cite{assran2023self}. It predicts the representations of various target blocks rather than reconstructing the masked input like MAE during pretraining. By predicting representations in the latent space, I-JEPA enhances the semantic quality of learned representations and boosts scalability and efficiency. 

Training a brain dynamics foundation model using a JEPA-like architecture might offer advantages over the MAE approach. However, the distinct spatiotemporal characteristics of fMRI data make direct application of the JEPA architecture suboptimal: \circledwhite{1} Positional embeddings in transformer play a crucial role by incorporating information about the order or position of tokens in the input data (\emph{e.g.}, the order of different words in a sentence or the locations of pixels in an image) \citep{vaswani2017attention}. However, there is no such natural "order" for different ROIs across the 3D brain volume in fMRI. BrainLM utilizes anatomical positions to label each ROI \citep{caro2023brainlm}, yet it does not account for brain functional parcellation, where nearby anatomical ROIs might exhibit rather different brain activation patterns represented by a lack of local coherence in fMRI data \citep{smith2013resting}. \circledwhite{2} I-JEPA employs a random multi-block selection of context and target. However, unlike images, fMRI presents complex patterns across both spatial and temporal domains. Given the smaller sample size and sparser information density in fMRI datasets compared to datasets like ImageNet \citep{deng2009imagenet}, learning in fMRI requires a stronger inductive bias. This would enhance the efficiency of training models by better capturing the underlying patterns specific to brain activity. Given the unique challenges presented by fMRI data, there is a pressing need to develop a functinal coordinate system and a tailored masking strategy for large-scale pretraining on fMRI data. 

\emph{These neglected yet crucial questions of developing a functional coordinate system and a masking strategy for large-scale pretraining with fMRI data, lie at the intersection of AI and neuroscience, highlighting important interdisciplinary challenges.}

To address these gaps, here we introduce \emph{Brain-JEPA}, a brain dynamics foundation model with the Joint-Embedding Predictive Architecture (JEPA). Instead of reconstructing masked inputs during pretraining, Brain-JEPA predicts abstract representations of sampled targets from the observation. We propose two innovative techniques to enhance model performance and address key questions in AI for neuroscience: First, \textbf{Brain Gradient Positioning} provides a brain functional coordinate system for positional embedding of brain functional parcellation (Section \ref{3.1}). Second, \textbf{Spatiotemporal Masking} offers a tailored masking strategy for the heterogeneous time-series patches inherent in fMRI (Section \ref{3.2}). Moreover, in downstream experiments, our proposed Brain-JEPA achieves state-of-the-art results in demographic prediction, disease diagnosis/prognosis, and trait prediction through fine-tuning. It also excels in off-the-shelf evaluations (\emph{e.g.}, linear probing), and shows superior generalizability across different ethnic groups. Brain-JEPA enhances brain activity analysis and deepens our understanding of critical AI-neuroscience questions related to constructing functional coordinate systems and developing spatiotemporal masking strategies.

\section{Related Work}

\textbf{Task-specific Models for fMRI (state-of-the-art).} SVR and MLP have been used in fMRI analysis, utilizing Pearson correlation matrices derived from fMRI time series as input \citep{drysdale2017resting,iidaka2015resting}. Deep learning models have substantially advanced fMRI analysis in recent years. BrainNetCNN \citep{kawahara2017brainnetcnn} introduces a convolutional neural network (CNN) with specialized convolutional filters tailored for brain network. BrainGNN \cite{li2021braingnn} utilizes ROI-aware graph neural networks (GNNs) to effectively harness functional brain network information, incorporating a pooling operator to highlight key ROIs. More recently, Brain network transformer (BNT) \citep{kan2022brain} employs transformer encoders to generate embeddings for ROIs based on Pearson correlation matrices, alongside a readout layer designed to identify clusters within the brain. Swift \cite{Swift} applies Swin Transformer architecture \cite{liu2021swin} to process brain functional data. As noted in Section \ref{intro}, these task-specific models have limited generalizability and adaptability across different tasks, and fail to utilize extensive unlabeled fMRI data.

\textbf{The fMRI Foundation Model.} BrainLM \citep{caro2023brainlm} stands out as the first fMRI foundation model, employing MAE for self-supervised pretraining of fMRI data. In this approach, fMRI time series are treated as images and patchified. The training goal is to reconstruct the masked patches of the time series. As outlined in Section \ref{intro}, BrainLM exhibits several limitations: 1) Direct reconstruction of masked input may not be suitable for inherently noisy data with low information density, such as fMRI. It complicates the differentiation between noise and signal, making it difficult to capture underlying patterns. 2) Generative architectures like MAE result in suboptimal performance in linear probing, a critical method for evaluating learned representations. 3) The absence of comparisons with state-of-the-art models and evaluations limited to Caucasian cohorts restricts its broader applicability. BrainMass \cite{BrainMass}, a concurrent work in large-scale self-supervised learning for neuroimaging, focuses on brain network analysis rather than brain dynamics, distinguishing it from our research.

\begin{figure}[]
    \centering
    \includegraphics[width=\columnwidth]{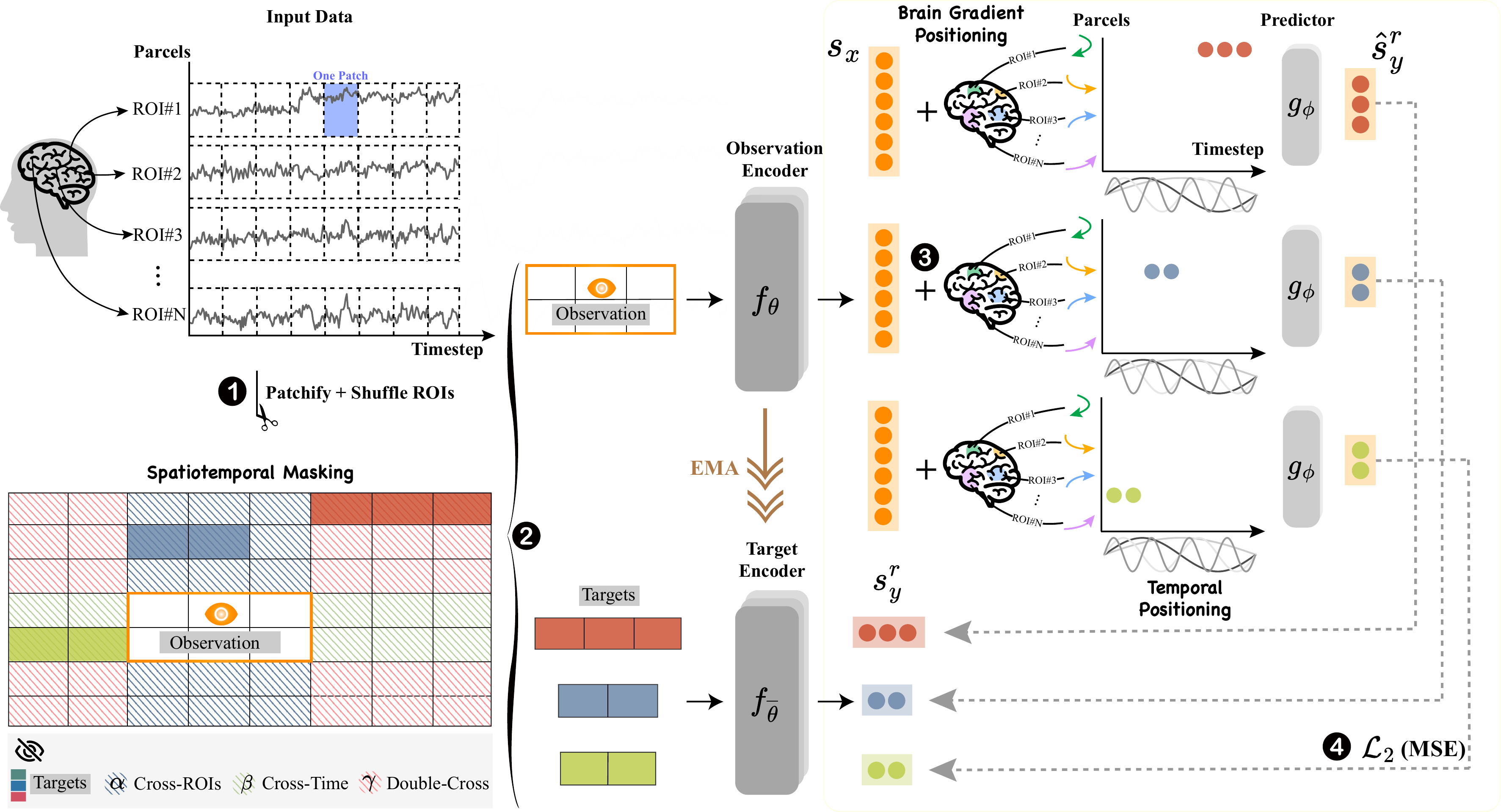}
    \caption{\textbf{Brain-JEPA.} With a Vision Transformer (ViT) as the observation encoder $f_{\theta}$, Brain-JEPA employs a single observation block to predict the representations of target blocks. \textbf{(1)} The input fMRI data is initially segmented into patches for subsequent processing. \textbf{(2)} Through Spatiotemporal Masking, the input data—excluding the observation block—is divided into three distinct regions: Cross-ROI ($\alpha$), Cross-Time ($\beta$), and Double-Cross ($\gamma$). The target blocks are sampled from different regions separately. \textbf{(3)} A narrower ViT, serving as the predictor $g_{\phi}$, takes the output $\boldsymbol{s}_x$ from $f_{\theta}$. It predicts the representations of a target block $\hat{\boldsymbol{s}}^{r}_y$ conditioned on positional embedding (brain gradient positioning for ROI locations and sine and cosine functions for temporal positioning). \textbf{(4)} These predicted representations align with those $\boldsymbol{s}^{r}_y$ from the target encoder $f_{\overline{\theta}}$, whose parameters are incrementally updated through an Exponential Moving Average (EMA) of the observation encoder's parameters.}
    \label{fig1}
\end{figure}

\section{Method}

In this section, we outline the methodology of Brain-JEPA. Instead of reconstructing masked patches of fMRI time series, Brain-JEPA operates in the latent space, as depicted in Figure \ref{fig1}. With the observation block excluded, the input data is divided into three non-overlapping regions: Cross-Time ($\alpha$), Cross-ROI ($\beta$), and Double-Cross ($\gamma$). This division forces the model to engage in forecasting time series, generalizing across unseen ROIs, and predicting time series for unseen ROIs. Section \ref{3.1} details the \textbf{Brain Gradient Positioning} we proposed, which encodes the functional relationships among different ROIs, serving as a brain functional coordinate system in the brain's functional organization. In Section \ref{3.2}, we introduce \textbf{Spatiotemporal Masking}, which injects a strong inductive bias during the masking process, leading to faster convergence during pretraining and superior performance in downstream tasks. 

\subsection{Brain Gradient Positioning}
\label{3.1}

\begin{wrapfigure}{r}{0.55\textwidth} 
  \centering
  \includegraphics[width=0.53\textwidth]{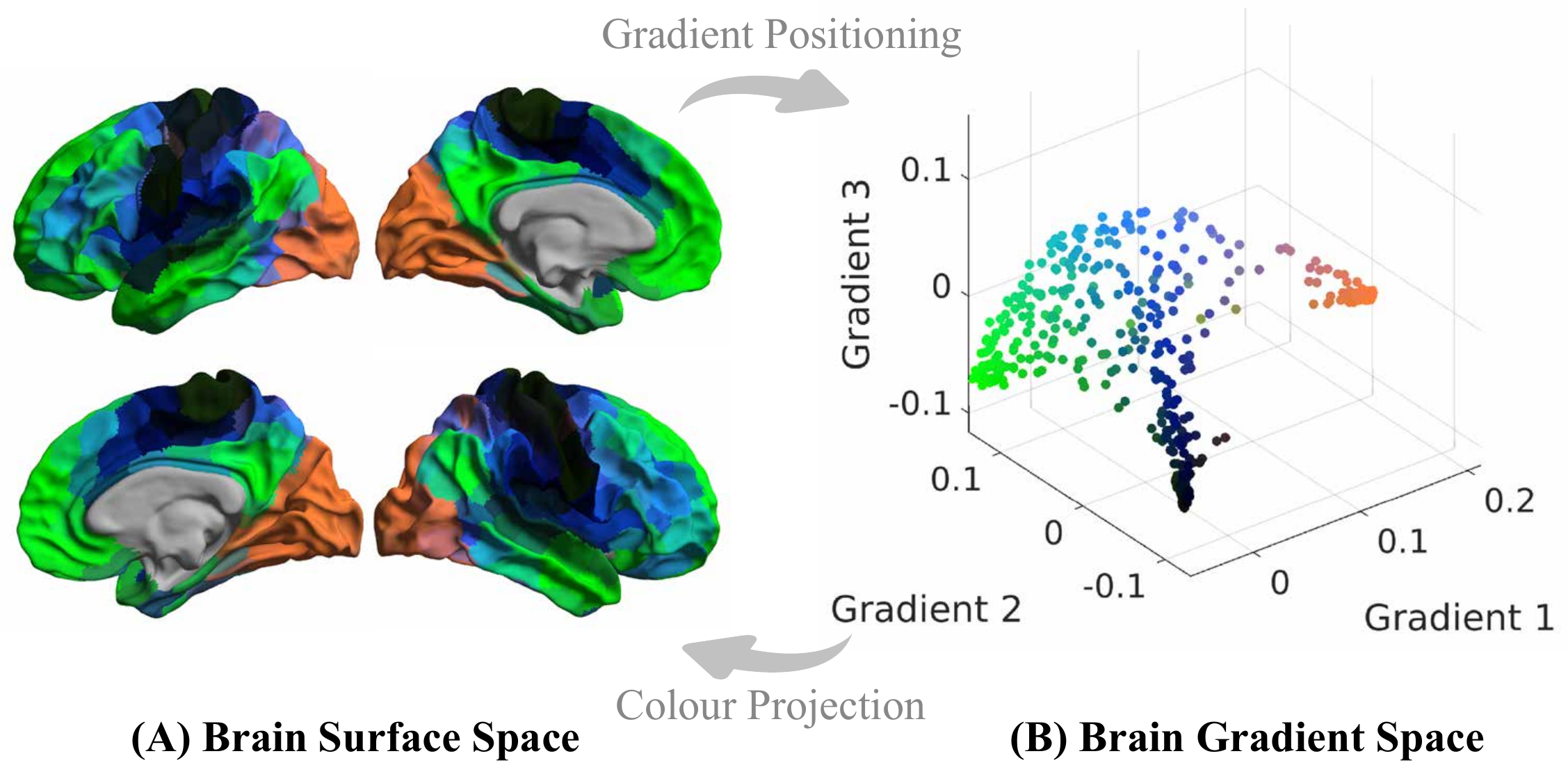} 
  \caption{\textbf{Brain gradient positioning.} Brain cortical regions are situated in the top 3 gradient axes and colored based on their positions. These colors are then projected back into the brain surface.}
  \label{figGrad2}
\end{wrapfigure}

We propose Brain Gradient Positioning, which provides a brain functional coordinates system based on the functional connectivity gradient. Positional embeddings are crucial in transformer architectures, as they encode information about the positions of tokens in a sequence. These embeddings can be implemented using fixed sine and cosine functions across various frequencies \citep{vaswani2017attention} or through learnable embeddings that adapt during training \citep{gehring2017convolutional}. However, the integration of positional information into fMRI time series has long been neglected. FMRI data, incorporating complex spatiotemporal information, requires separate consideration of its temporal and spatial domains. The temporal domain, representing timesteps during scanning, is well-suited for conventional sine and cosine positional embeddings, as the time series in each ROI is sequentially ordered by time. However, this method is not appropriate for the spatial domain, where ROIs across brain volumes lack a simple, inherent order, making sine and cosine embeddings unsuitable for capturing spatial relationships. Anatomical locations of ROIs offer an alternative to sine and cosine functions \cite{caro2023brainlm} but fall short in capturing functional parcellation. Spatially adjacent ROIs can exhibit significantly different brain activation patterns, reflecting the inherent lack of local coherence in fMRI data \cite{smith2013resting}.

The functional connectivity gradient is a continuous measure that captures the functional relations among different ROIs. Each attribute in the gradient represents an axis in the latent space of brain regions and networks. The relative distance between different ROIs indicates the similarity in their connectivity (\emph{i.e.}, shorter distance means higher similarity in connectivity). The concept of a spatial gradient as conceptualized by Mesulam in 1998 entailed a synaptic hierarchy that supports cognitive processes \cite{Mesulam1998}. Recent studies have built upon this concept, revealing that brain networks in adult humans and macaques exhibit linear distributions across different gradient axes \cite{Margulies2016grad}. Using this methodology, it has been shown that these gradients reflect the functional changes related to age \cite{Zhou2023JAACAP,Lariviere2020gradnatal,Nguyen2023grad,Dong2021gradchild,Bethlehem2020gradlife}, cognition \cite{Bethlehem2020gradlife,Wang2020gradcog} and brain diseases \cite{Nguyen2023grad,Hong2019gradautism,He2023gradAD}.  These gradients together provide a framework to assess the relationship between brain regions based on their relative positioning across different gradient axes. 

Before deriving the gradients, we first calculate a non-negative affinity matrix $\boldsymbol{A}(i,j)$ (a graph Laplacian) as follows:
\vspace{-0.02cm}
\begin{equation}
\boldsymbol{A}(i,j) = 1 - \frac{1}{\pi} \cos^{-1}(\frac{{\mathbf{c}_i \mathbf{c}_j^T}}{{\left\| \mathbf{c}_i \right\| \left\| \mathbf{c}_j \right\|}})
\end{equation}
\vspace{-0.02cm}
where $\boldsymbol{c}_i$ and $\boldsymbol{c}_j$ represents the features (functional connectivity) across the ROI $i$ and $j$, respectively. Gradients are then derived using diffusion map \cite{Coifman2005DM,vos2020brainspace}, a nonlinear dimension reduction method used to identify the underlying manifold structure of the data. We can obtain the diffusion matrix $\boldsymbol{L}_{\delta}$ and the diffusion operator $\boldsymbol{M}_\delta$ from $\boldsymbol{A}$ as follows:
\vspace{-0.02cm}
\begin{equation}
       \boldsymbol{M}_\delta = \boldsymbol{D}^{ - 1}\boldsymbol{L}_\delta, \ \boldsymbol{L}_\delta = \boldsymbol{D}^{ - \frac{1}{\delta} } \boldsymbol{A} \boldsymbol{D}^{ - \frac{1}{\delta} } 
\end{equation}
\vspace{-0.02cm}
where $\boldsymbol{D}$ is the degree matrix  of $\boldsymbol{A}$. Here $\delta$ is set to 0.5 to maintain the global relations between ROIs in the embedding space.

Finally, we can compute the eigenvectors and eigenvalues of $\boldsymbol{M}_{\delta}$, and stack the column vectors to formulate the diffusion map \( \boldsymbol{\Phi}_t \in \mathbb{R}^{n\times m}\) ($n$ ROIs in total, with $m$ gradients for each) at time \( t \) and the gradient matrix $\boldsymbol{G}$ with the same dimension:
\vspace{-0.03cm}
\begin{equation}
    \boldsymbol{\Phi}_t = [ \lambda_1^t \psi_1, \lambda_2^t \psi_2, \ldots, \lambda_m^t \psi_m ], \ \boldsymbol{G} = [ \psi_1, \psi_2, \ldots, \psi_m ]
\end{equation}
\vspace{-0.02cm}
where \( \lambda_k \) are the eigenvalues and \( \psi_k \) are the corresponding eigenvectors (gradients) of the graph Laplacian. The parameter \( t \)  represents the diffusion time, which controls the scale of the diffusion process. Here we estimated the eigenvalues \( \lambda_k \) at time \( t \) by dividing it by \(1- \lambda_k \) to enhance robustness against noisy eigenvalues.

In Brain-JEPA, we leverage $\boldsymbol{G}$ as the spatial positioning of ROIs. Specifically, the gradient $\boldsymbol{G}\in \mathbb{R}^{n\times m}$ is transformed into $\hat{\boldsymbol{G}}\in \mathbb{R}^{n\times d/2}$ through a trainable linear layer, where $d$ represents the embedding dimension of the ViT backbone. The predefined temporal positioning $\boldsymbol{T}\in \mathbb{R}^{n\times d/2}$ is obtained using sine and cosine functions \citep{NIPS2017_3f5ee243}. The final positional embedding can then be formulated as $\boldsymbol{P}=[\boldsymbol{T}, \hat{\boldsymbol{G}}] \in \mathbb{R}^{n\times d}$.  Figure \ref{figGrad2} provides a visualization of the top 3 gradients in Euclidean space with each ROI color coded by their locations. As shown, the brain gradient positioning reflects functional network architecture, such as the \textbf{\textcolor{soma}{somatomotor}}, \textbf{\textcolor{def}{default mode}} and \textbf{\textcolor{vis}{visual}} networks, consistent with previous literature \cite{Margulies2016grad,Bethlehem2020gradlife}.

\subsection{Spatiotemporal Masking}
\label{3.2}

\textbf{Observation.} Brain-JEPA aims to predict representations of multiple target blocks based on the representation of a single observation block. For an input fMRI time series, the temporal signal for each parcel is divided into patches after shuffling ROIs, each containing $p$ time points (dash boxes in Figure \ref{fig1}). The observation block $\boldsymbol{x}$ is obtained by randomly sampling a block within the range \{$\eta^{\text{o}}_R$,\ $\eta^{\text{o}}_T$\}. $\eta^{\text{o}}_R$ specifies the range ratio along the ROI dimension, and $\eta^{\text{o}}_T$ pertains to the timestep patches (10 in total). Subsequently, $\boldsymbol{x}$ is fed through the observation encoder $f_{\theta}$, generating a corresponding patch-level representation $\boldsymbol{s}_x$:
\vspace{-0.02cm}
\begin{equation}
    \boldsymbol{s}_x=\{\boldsymbol{s}_{x_j}\}_{j\in \mathcal{B}_x}
\end{equation}

where $\mathcal{B}_x$ represents the mask associated with the observation block $\boldsymbol{x}$, $\boldsymbol{s}_{x_j}$ is the representation of the $j^{\text{th}}$ patch.

\textbf{Targets.} Given a single observation, the model is trained to predict other parts of the fMRI within the latent space. Random sampling of targets like MAE \citep{He2022MAE} might allow the model to learn shortcuts (\emph{e.g.}, interpolation of time series) or rely heavily on simpler, more frequent patterns in the data, which could limit its generalizability. It is crucial to recognize that patches in fMRI vary spatially depending on the positions in their brain functional organization, and temporally regarding brain states and task conditions. The nonlinear relationship among brain networks further complicates the interactions between different brain patches. 

As shown in Figure \ref{fig1}, we categorize the remaining parts (with the observation excluded) into three distinct and non-overlapping regions: Cross-ROI ($\alpha$), Cross-Time ($\beta$), and Double-Cross ($\gamma$). For targets in the $\alpha$ and $\beta$ regions, the model should generalize the observation across different ROIs spatially or timesteps temporally. For targets in the $\gamma$ regions, which are the most challenging, the model should generalize to unseen ROIs at unencountered timesteps. We randomly sample $K$ blocks from each of the three types of regions as targets, forcing the model to handle a variety of prediction tasks with a stronger inductive bias. We denote the mask corresponding of the region $r$ ($r\in \{\alpha, \beta, \gamma\}$) as $\mathcal{B}_y^r$.

\textbf{Overlapped sampling.} It has been shown in \cite{assran2023self} that a sufficiently large dynamic range of masking ratio could benefit pretraining. To effectively adjust the observation-to-input ratio during pretraining, we implement an overlapped sampling strategy that allows for a flexible, rather than fixed, ratio. When sampling the target block $\boldsymbol{s}_{y}^{r}$ from region $r$, for $r=\alpha \ \text{or} \ \beta$, we sample the target from the union of the observation mask and region $r$ mask; while for $r=\gamma$, we directly sample the target from the $\gamma$ region mask. Formally, the overlapped sampling strategy is defined as:
\vspace{-0.001cm}
\begin{equation}
    \boldsymbol{s}_{y}^{\alpha} \sim \mathcal{B}_x \cup \mathcal{B}_y^{\alpha}, \ \boldsymbol{s}_{y}^{\beta} \sim \mathcal{B}_x \cup \mathcal{B}_y^{\beta}, \ \boldsymbol{s}_{y}^{\gamma} \sim \mathcal{B}_y^{\gamma}
\end{equation}
\vspace{-0.001cm}
Afterwards, part of the observation region might overlap with some $\alpha$ and $\beta$ targets. We remove any ROIs in the observation that overlap with the $\alpha$ targets. Additionally, we eliminate all timesteps for ROIs that show overlap with the $\beta$ targets. Refer to Table \ref{tab6} for the block sizes.

\textbf{Training.} Given the output $\boldsymbol{s}_x$ from the observation encoder $f_{\theta}$, the predictor $g_{\phi}$ is trained to predict the three kinds of targets $\boldsymbol{s}_{y}^{r}$ conditioned on the positional embedding $\boldsymbol{P}$ (Figure \ref{fig1}). The training loss $\mathcal{L}$ is the average $L_2$ distance between $\boldsymbol{s}_{y}^{r}$ and its corresponding prediction:
\vspace{-0.001cm}
\begin{equation}
    \mathcal{L}=\frac{1}{3K}\sum_{r} \left\|\hat{\boldsymbol{s}}_y^r - \boldsymbol{s}_{y}^{r}\right\|_2^2, \ \hat{\boldsymbol{s}}_y^r = g_{\phi}(\boldsymbol{s}_x|\boldsymbol{P})
\end{equation}
\section{Experiments}

\subsection{Datasets}
\label{dataset}

We leveraged the large-scale public dataset - UK Biobank (\href{https://www.ukbiobank.ac.uk/}{UKB}) \citep{miller2016multimodal} for the self-supervised pretraining of Brain-JEPA. It includes resting-state fMRI recordings with medical records from 40,162 participants aged 44 to 83. Multi-site recordings were acquired with the temporal resolution of 0.735s. We allocated 80\% of this dataset for pretraining (of which we calculated the group-level gradients as well), with the 20\% held-out for downstream evaluation (internal tasks of age and sex prediction). 

We used three datasets for external evaluation: \href{https://www.humanconnectome.org/study/hcp-lifespan-aging}{HCP-Aging}, as a segment of the public Human Connectome Project (HCP) \citep{elam2021human}, includes resting-state fMRI from 656 healthy elderly participants. It was used to predict traits (Neuroticism and Flanker score) and demographics (age and sex). The Alzheimer’s Disease Neuroimaging Initiative (\href{https://adni.loni.usc.edu/data-samples/access-data/}{ADNI}) \citep{jack2008alzheimer} was used for the early diagnosis and prognosis of neurodegenerative diseases, with fMRI from 189 participants for normal control (NC) \textit{v.s.} mild cognitive impairment (MCI) classification, and 100 cognitively normal participants for amyloid positive v.s. negative classification. Moreover, to assess generalizability across different ethnic groups and real-world clinical applications, we included resting-state fMRI of Asian participants recruited by Memory, Ageing and Cognition Centre \href{https://medicine.nus.edu.sg/macc/}{(MACC)}, with 539 participants for NC v.s. MCI classification. More details of the downstream tasks performed can be found in the Appendix \ref{A}.

All fMRI data was parcellated into $n=450$ ROIs, using Schaefer-400 \citep{schaefer2018local} for cortical regions and Tian-Scale III \citep{tian2020topographic} for subcortical regions. Robust scaling was implemented by subtracting the median and dividing by the interquartile range, calculated across participants for each ROI \citep{caro2023brainlm}. Our default input size is 160 timesteps for each of the 450 ROIs (\emph{i.e.}, 450×160). UKB and HCP-Aging used multi-band acquisition with a high temporal resolution (TR $\approx 0.7$ seconds), while ADNI and MACC used single-band acquisition with a lower resolution (TR $\approx 2$ seconds). To ensure consistency across datasets, we standardized the temporal resolution by downsampling the multi-band data using a temporal stride of 3, aligning the TR of all datasets to approximately 2 seconds. During the fine-tuning and linear probing stage, all the downstream datasets were divided into a 6:2:2 ratio for training, validation, and testing.

\subsection{Implementation details}
\label{implementation}

For Brain-JEPA pretraining, we utilized ViT architectures for the observation encoder, target encoder, and predictor. We employed FlashAttention \cite{dao2022flashattention,dao2023flashattention2} in our self-attention implementation to improve computational efficiency and reduce memory usage. Balancing the trade-off between data quantity and the model complexity, we experimented with ViT-Small (ViT-S) (22M), ViT-Base (ViT-B) (86M), and ViT-Large (ViT-L) (307M) for the observation encoder. For predictor, it is designed as a lightweight (narrow) ViT. Specifically, the predictor has the same architecture as the corresponding observation encoder, differing only in embedding dimension and depth. For the ViT-S and ViT-B observation encoders, the predictor has a depth of 6 and embedding dimensions of 192 and 384, respectively. The ViT-L observation encoder uses a predictor with a depth of 12 and an embedding dimension of 384. Brain-JEPA is pretrained without a [cls] token. For evaluation, we used the target encoder and average pooled its output to generate a global fMRI representation. The main results in Section \ref{4.3}, along with the analysis in Section \ref{sec4.5}, \ref{sec4.6} and \ref{sec4.7} were all based on ViT-B pre-trained for 300 epochs. Refer to Appendix \ref{B} for optimization and masking details.

\subsection{Main results}
\label{4.3}

\begin{table}[t]
\centering
\small
\caption{Internal tasks of age and sex prediction on UKB 20\% held-out. 
 The mean (standard deviation) of Mean Squared Error (MSE), Pearson Correlation ($\rho$), and/or Accuracy (ACC), F1 score across 5 independent runs is reported. $\uparrow$: the higher, the better; $\downarrow$: the lower, the better. The best results are in \textbf{bold}, with \textbf{*} denoting significant improvement over previous approaches ($p<0.05$).
}
\renewcommand{\arraystretch}{1.5}
\resizebox{0.59\textwidth}{!}{
\begin{tabular}{lcccccc}
\toprule
\multirow{2.5}{*}{\large Methods} & \multicolumn{2}{c}{\large Age} & \multicolumn{2}{c}{\large Sex} \\
\cmidrule(lr){2-3} \cmidrule(lr){4-5}
& \large MSE $\downarrow$ & \large $\rho$ $\uparrow$ & \large ACC(\%) $\uparrow$ & \large F1(\%) $\uparrow$ \\
\midrule
\large BrainNetCNN \cite{kawahara2017brainnetcnn} & 0.985 (0.027) & 0.225 (0.015) & 77.86 (0.98) & 78.17 (0.86) \\
\large BrainGNN \cite{li2021braingnn} & 0.931 (0.038) & 0.332 (0.015) & 77.31 (0.33) & 79.23 (0.31) \\
\large BNT \cite{kan2022brain} & 0.863 (0.031) & 0.447 (0.017) & 80.78 (0.40) & 82.42 (0.36) \\
\large TFS\textsuperscript{†} & 0.812 (0.023) & 0.487 (0.011) & 82.60 (0.59) & 83.00 (0.01) \\
\large BrainLM \cite{caro2023brainlm} & 0.612 (0.041) & 0.632 (0.020) & 86.47 (0.74) & 86.84 (0.43)\\ 
\rowcolor{lightgray}\textbf{\large Brain-JEPA} & \textbf{0.501*} (0.034)  & \textbf{0.718*} (0.021)  & \textbf{88.17*} (0.06) & \textbf{88.58*} (0.11) \\
\bottomrule
\end{tabular}}
\begin{tablenotes}
   \item[*] \hspace{2.5cm}\scriptsize \textsuperscript{†} Trained-From-Scratched.
\end{tablenotes}
\label{tab1}

\end{table}

\begin{table}[t]
  \caption{External tasks of demographics and trait prediction on HCP-Aging.}
  \label{tab:External tasks on HCP.}
  \centering
\renewcommand{\arraystretch}{1.5}
\resizebox{\textwidth}{!}{
\begin{tabular}{lcccccccc}
\toprule
\multirow{2}{*}{\large Methods} & \multicolumn{2}{c}{\large Age}     & \multicolumn{2}{c}{\large Sex}     & \multicolumn{2}{c}{\large Neuroticism} & \multicolumn{2}{c}{\large Flanker}     \\ 
\cmidrule(lr){2-3} \cmidrule(lr){4-5} \cmidrule(lr){6-7} \cmidrule(lr){8-9}
                         & \large MSE $\downarrow$          &  \large $\rho$ $\uparrow$            & \large ACC (\%) $\uparrow$     & \large F1 (\%) $\uparrow$      & \large MSE $\downarrow$            &  \large  $\rho$ $\uparrow$             & \large MSE $\downarrow$            & \large  $\rho$ $\uparrow$              \\ \hline
\large BrainNetCNN \cite{kawahara2017brainnetcnn}             & 0.462 (.017) & 0.611 (.023) & 71.16 (0.88) & 72.23 (0.92) & 1.201 (.097) & 0.096 (.006) & 1.045 (.036) & 0.201 (.018) \\
\large BrainGNN \cite{li2021braingnn}                & 0.423 (.015) & 0.672 (.024) & 72.7 (0.54)   & 74.09 (0.67) & 1.183(.096)    & 0.098 (.007)   & 0.982 (.043)   & 0.309 (.062)   \\
\large BNT \cite{kan2022brain}                      & 0.414 (.035) & 0.731 (.057) & 72.41 (1.09) & 73.68 (1.11) & 1.199 (.091)   & 0.101 (.005)   & 0.997 (.037)   & 0.307 (.026)   \\
\large BrainLM \cite{caro2023brainlm}                 & 0.331 (.018) & 0.832 (.028) & 74.39 (1.55) & 77.51 (1.13) & 0.942 (.082)   & 0.231 (.012)   & \textbf{0.971 (.054)}   & 0.318 (.048)   \\
\rowcolor{lightgray}\textbf{\large Brain-JEPA}               & \textbf{0.298 (.017)} & \textbf{0.844 (.030)} & \textbf{81.52* (1.03)} & \textbf{84.26* (0.82)} & \textbf{0.897* (.055)}  & \textbf{0.307* (.006)}  & 0.972 (.038)   & \textbf{0.406* (.027)}  \\  \bottomrule
\end{tabular}}
\label{tab2}

\end{table}

{
\begin{table}[t]
\centering
\small
\caption{External tasks of brain disease diagnosis and prognosis on ADNI and MACC.}
\renewcommand{\arraystretch}{1.5}
\resizebox{0.83\textwidth}{!}{
\begin{tabular}{lcccccc}
\toprule
\multirow{2.5}{*}{\large Methods} & \multicolumn{2}{c}{\large NC/MCI} & \multicolumn{2}{c}{\large Amyloid $a\beta+$ve/$-$ve} & \multicolumn{2}{c}{\large NC/MCI (Asian)}\\
\cmidrule(lr){2-3} \cmidrule(lr){4-5} \cmidrule(lr){6-7}
& \large ACC(\%) $\uparrow$ & \large F1(\%) $\uparrow$ & \large ACC(\%) $\uparrow$ & \large F1(\%) $\uparrow$ & \large ACC(\%) $\uparrow$ & \large F1(\%) $\uparrow$\\
\midrule
\large BrainNetCNN \cite{kawahara2017brainnetcnn} & 60.00 (3.51) & 64.72 (3.18) & 59.00 (2.00) & 59.43 (1.14) & 57.32 (4.45) & 53.92 (4.25)\\
\large BrainGNN \cite{li2021braingnn} & 67.40 (2.93) & 71.42 (2.87) & 57.00 (4.00) & 62.61 (3.48) & 59.79 (2.35) & 55.69 (2.29)\\
\large BNT \cite{kan2022brain} & \textbf{78.90} (4.12) & 83.14 (3.58) & 62.00 (2.45) & 59.53 (0.58) & 62.06 (3.88) & 60.45 (4.52)\\
\large BrainLM \cite{caro2023brainlm} & 75.79 (1.05) & 85.66 (1.27) & 67.00 (7.48) & 68.82 (8.48) & 61.65 (3.35) & 60.26 (3.03)\\
\rowcolor{lightgray}\textbf{\large Brain-JEPA} & 76.84 (1.05)  & \textbf{86.32} (0.54)  & \textbf{71.00*} (4.90) & \textbf{75.97*} (3.93) & \textbf{65.98*} (2.84) & \textbf{64.67*} (2.61)\\
\bottomrule
\end{tabular}}
\label{tab3}
\end{table}
}

Table \ref{tab1}, \ref{tab2}, and \ref{tab3} compare Brain-JEPA with the existing deep learning models for fMRI analysis and foundation model BrainLM. We select the three deep learning baselines because they not only represent the previous state-of-the-art in fMRI analysis but also exemplify diverse model types: convolutional neural network (CNN)-based BrainNetCNN \cite{kawahara2017brainnetcnn}, graph neural network (GNN)-based BrainGNN \cite{li2021braingnn}, and transformer-based BNT \cite{kan2022brain}. For a fair comparison, both Brain-JEPA and BrainLM utilized a ViT-B backbone and were fine-tuned for downstream tasks (Section \ref{sec4.4} will discuss performance scaling with different model sizes, and Section \ref{sec4.5} will examine linear probing comparisons between the two models). BrainLM utilized [cls] token for downstream evaluation.

The results show that Brain-JEPA achieves state-of-the-art performance in various downstream tasks on both the unseen data from the same pretrained cohort and other independent datasets. Brain-JEPA effectively captures fundamental demographic information such as age and sex, cognitive/personality variance (Neuroticism and Flanker), and disease-related patterns for neurodegenerative diseases. Notably, Brain-JEPA demonstrates superior performance in classifying NC/MCI in Asian ethnic groups — one of the most challenging tasks for early diagnosis and prognosis of Alzheimer's Disease (AD) — even though it was trained exclusively on the Caucasian cohort. Please refer to \ref{D.1} for additional results on more datasets and comparisons with more baselines.

\subsection{Performance scaling}
\label{sec4.4}

\begin{figure}[t]
    \centering
    \includegraphics[width=0.8\columnwidth]{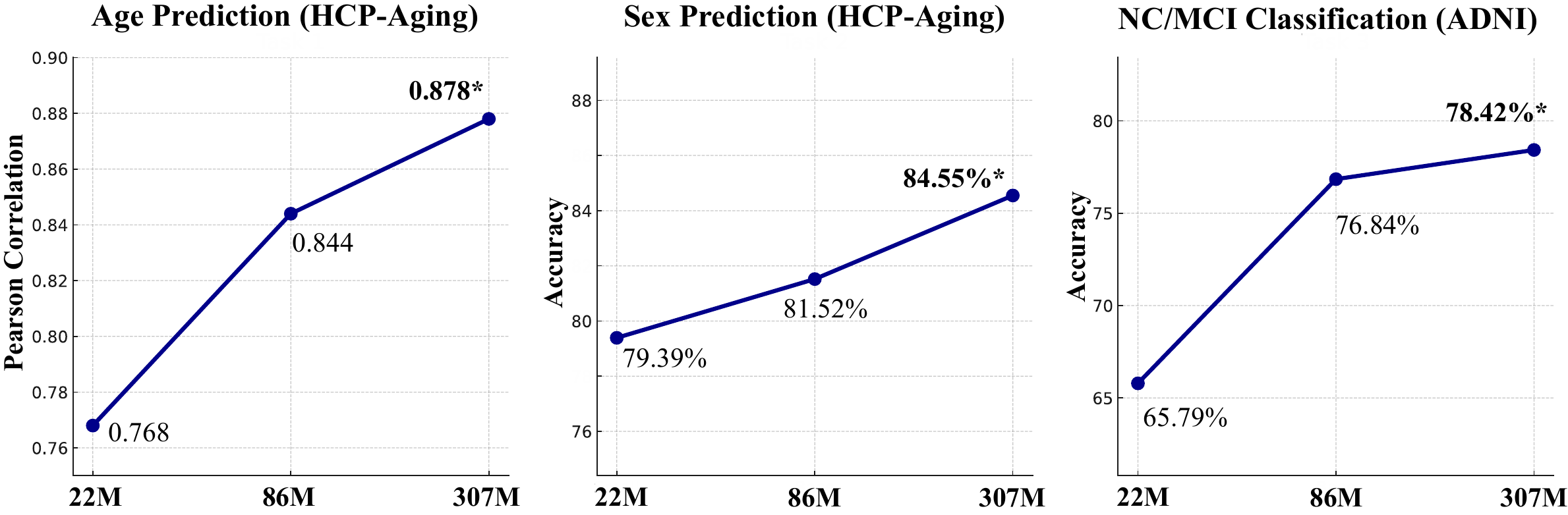}
    \caption{Performance scaling of the model sizes.}
    \label{scaling}
\end{figure}

Figure \ref{scaling} presents the performance of Brain-JEPA across various model sizes, using ViT-S, ViT-B, and ViT-L as backbones. The results demonstrate that the larger model configuration consistently achieves better performance. Specifically, there is a clear trend of increasing accuracy/correlation with larger models, with Brain-JEPA using ViT-L consistently achieving the best performance. We also studies the scaling property with respect to dataset size, please refer to \ref{D.3} for additional results.

\subsection{Linear probing}
\label{sec4.5}

\begin{figure}[t]
    \centering
    \includegraphics[width=0.8\columnwidth]{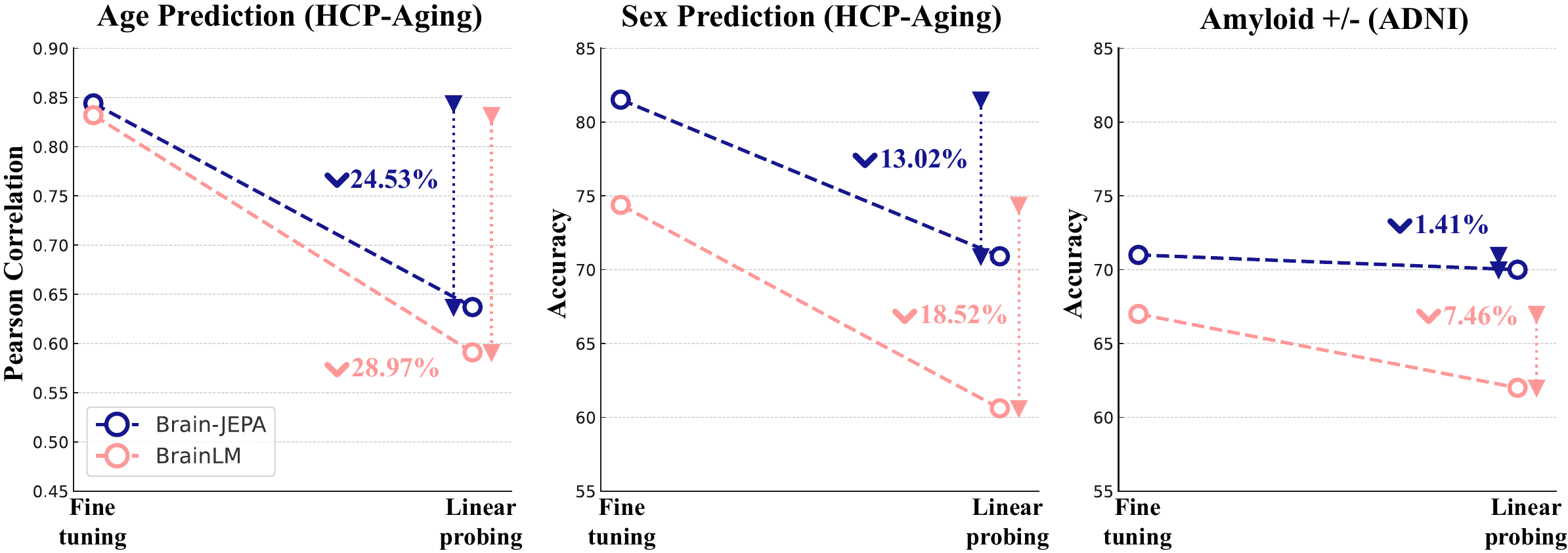}
    \caption{{Fine-tuning \textit{v.s.} linear probing.}}
    \label{lp}
\end{figure}

BrainLM initially showcases its performance improvements through fine-tuning, complemented by an attached MLP \citep{caro2023brainlm}. However, to effectively assess the representations learned during pretraining, off-the-shelf evaluations such as linear probing are essential. As depicted in Figure \ref{lp}, Brain-JEPA consistently outperforms BrainLM in linear probing and exhibits a smaller performance decline from fine-tuning to linear probing. This highlights the robustness and higher level of abstraction in the representations learned by Brain-JEPA.

\begin{wrapfigure}{r}{0.45\textwidth} 
\vspace{-2.5em}
  \centering
  \includegraphics[width=0.43\textwidth]{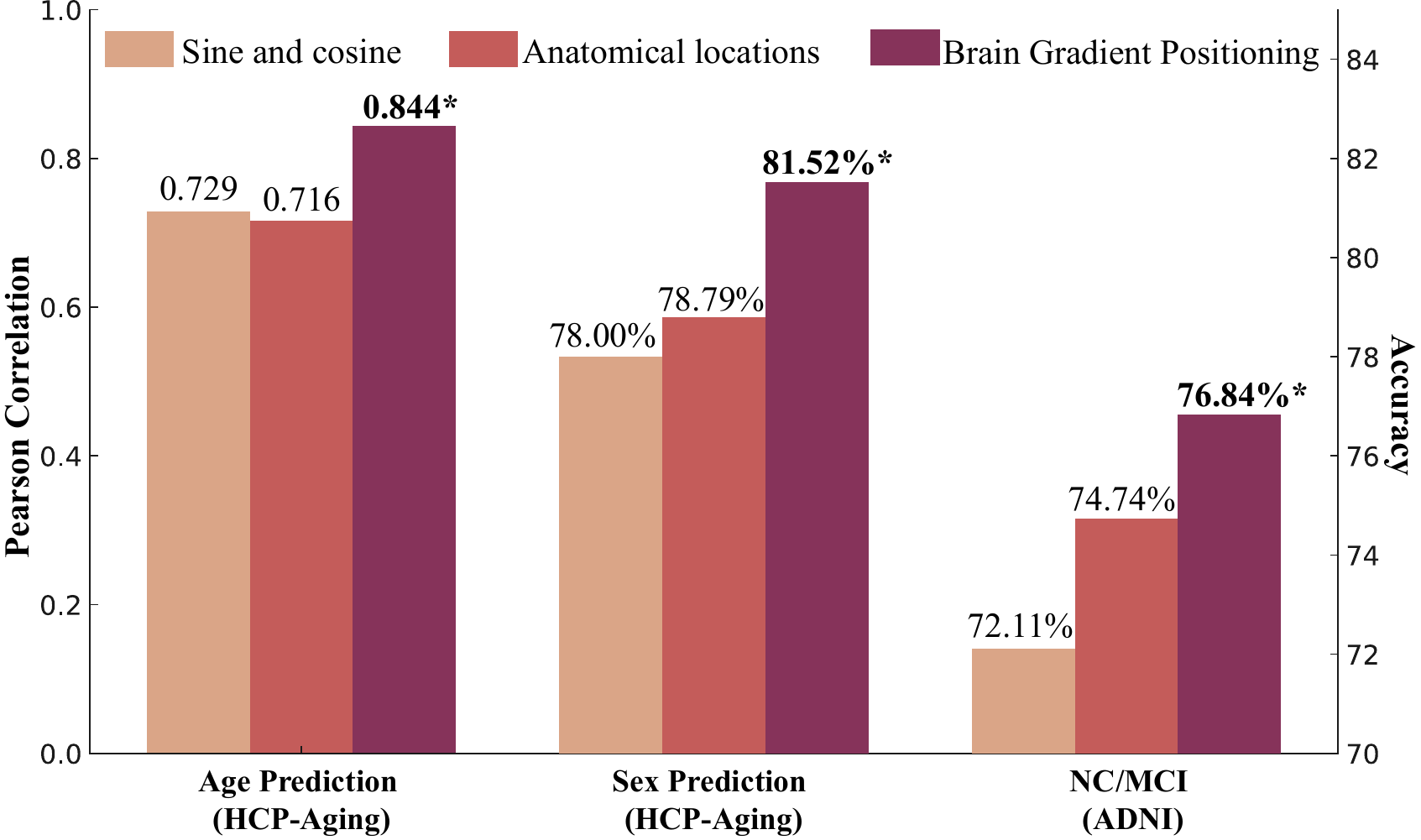} 
  \caption{Comparisons of spatial positional embedding (For the first task, refer to the left $y$ axis for the Pearson's Correlation, with the right $y$ axis accuracy for the last two tasks).}
  \label{position}
\vspace{-0.5cm}
\end{wrapfigure}

\subsection{Ablation study}
\label{sec4.6}
\begin{figure}[t]
    \centering
    \includegraphics[width=0.8\columnwidth]{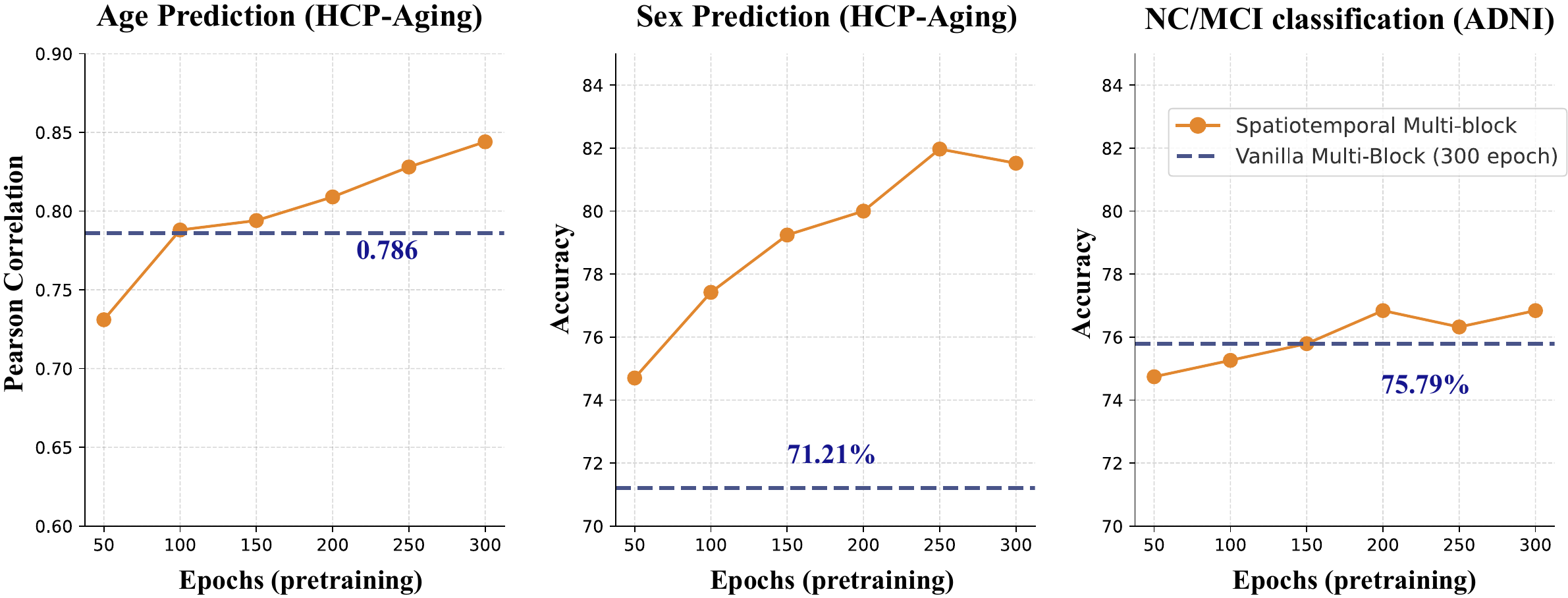}
    \caption{Comparisons of masking strategies.}
    \label{block}
\end{figure}

We first compared Brain-JEPA with its ablated versions, employing sine and cosine functions \citep{NIPS2017_3f5ee243} and anatomical locations \citep{caro2023brainlm} for ROI spatial positioning, as shown in Figure \ref{position}. Brain Gradient Positioning demonstrates superior performance over these two baseline methods. It indicates that Brain Gradient Positioning facilitates natural and accurate placement of brain functional parcellations, enhancing the learning of brain dynamics. Next, we assessed the effectiveness of our proposed Spatiotemporal Masking by comparing Brain-JEPA, pretrained over various numbers of epochs, to its ablated counterpart that utilizes standard multi-block sampling of targets \cite{assran2023self}. This comparison, illustrated in Figure \ref{block}, highlights that not only does our proposed masking technique yield superior performance, but it also introduces a stronger inductive bias leading to a more efficient pretraining. Notably, Brain-JEPA achieves or even surpasses the peak performance of the ablated version, which was pretrained for 300 epochs, with significantly fewer epochs—only 100, 200, and 50 respectively. For more ablation results regarding architectures and the number of gradient components, please refer to \ref{D.2}.

\subsection{Interpretation}
\label{sec4.7}

\begin{figure}[!h]
    \centering
    \includegraphics[width=\columnwidth]{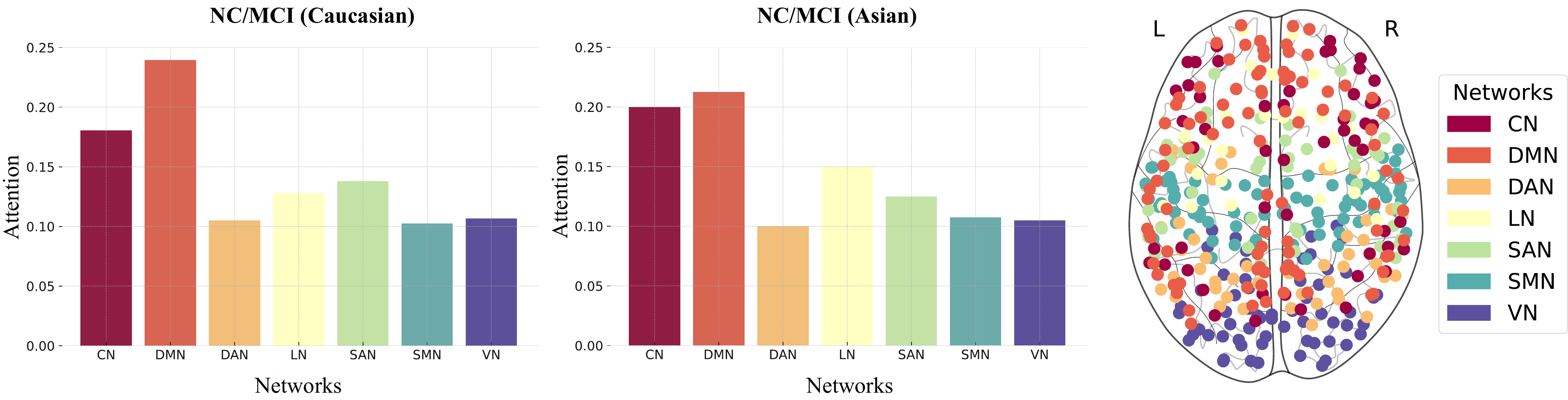}
    \caption{Attention across different brain networks for NC/MCI classification.}
    \label{attention}
\end{figure}

With the Schaefer functional atlas \cite{schaefer2018local}, the brain network is categorized into seven distinct sub-networks: the control network (CN), the default mode network (DMN), the dorsal attention network (DAN), the limbic network (LN), the salience ventral attention network (SAN), the somatomotor network (SMN), and the visual network (VN). To assess whether Brain-JEPA has captured the brain functional organization, we calculate the network-level attention for NC/MCI classification. For each ROI, we first average the self-attention across its 10 patches. Next, we average the values of the ROIs within each sub-network and normalize them to obtain the network-level attention distribution. As shown in Figure \ref{attention}, we found consistent patterns across both Caucasian and Asian ethnic groups, with the model highlighting the critical roles of the DMN, CN, SAN, and LN in cognitive impairment, consistent with previous literature \cite{talwar2021systematic,sheline2013resting,brier2014network}.

\section{Conclusion}

In this study, we developed Brain-JEPA, a brain dynamics foundation model based on the Joint-Embedding Predictive Architecture (JEPA). Brain-JEPA predicts abstract representations of sampled targets from observations during the pretraining stage. Utilizing Brain Gradient Positioning, Brain-JEPA encodes brain functional organization more naturally and accurately. With Spatiotemporal Masking, it effectively handles heterogeneous patches in fMRI time series. Brain-JEPA fosters generalizable and highly abstract representations of fMRI, achieving state-of-the-art performance across various tasks, including demographic prediction, trait prediction, and disease diagnosis and prognosis across different cohorts and ethnic groups. Our study provides new insights into applying large-scale self-supervised learning to brain activity modelling and contributes to addressing key questions in AI for neuroscience.

\section{Limitation and future work}
\label{limitation}

We acknowledge several limitations in our study, which also serve as inspirations for future research: 1) Larger models: Due to limited computing resources, we have not tested larger models like ViT-H. We expect that larger models could further improve performance. 2) More diverse datasets: A more diverse brain recording dataset for pretraining, including different ethnicity cohorts collected from various sites, scanning protocols, behavioral tasks, and disease groups, could enhance the generalizability and robustness of the representations learned by the model. 3) Fine-grained interpretation: More thorough interpretation can be achieved through the attention mechanism, such as comparing cortical and subcortical regions, identifying salient ROIs and critical timesteps. This would enable more nuanced and complex spatiotemporal interpretations. 4) Multi-modal integration: Brain-JEPA sets a potential foundation for integrating multimodal brain activity data such as MEG and EEG or even brain structure data like T1-weighted MRI. The integration could enhance our understanding of brain structure, function, and their links to human behavior and mental disorders. Please refer to Appendix \ref{C} for the broader impact of Brain-JEPA.

\bibliographystyle{unsrtnat}
\bibliography{ref}


\newpage
\appendix

\section{Task Details}
\label{A}

\subsection{Neuroticism}

Neuroticism is a personality trait linked to negative emotions and is one of the Big Five personality traits. People who score high in neuroticism tend to experience negative feelings more frequently than others \cite{thompson2008development}. The HCP uses the 60-question version of the NEO-FFI Short Form (ages 16+) questionnaire, which provides a quick, reliable, and accurate assessment of the Big Five personality traits: neuroticism, extraversion, openness, agreeableness, and conscientiousness. More detailed information can be found in the \href{https://www.humanconnectome.org/study/hcp-lifespan-aging/document/hcp-aging-20-release}{Lifespan HCP 2.0} Data Release Appendix 2: Details and References for Behavioral \& Clinical Instruments.

\subsection{Flanker}

The Flanker task is designed to assess both attention and inhibitory control in participants \cite{eriksen1974effects}. It involves the participant focusing on a central stimulus while ignoring adjacent stimuli, which are either fish for ages 3-7 or arrows for ages 8-85. Sometimes the central stimulus points in the same direction as the flanking stimuli (congruent) and sometimes in the opposite direction (incongruent). For participants aged 8-85, the task consists of twenty trials and takes about three minutes to complete. More detailed information can be found in the \href{https://www.humanconnectome.org/study/hcp-lifespan-aging/document/hcp-aging-20-release}{Lifespan HCP 2.0} Data Release Appendix 2: Details and References for Behavioral \& Clinical Instruments.

\subsection{NC/MCI}

For the \href{https://adni.loni.usc.edu/data-samples/access-data/}{ADNI} dataset \citep{jack2008alzheimer}, the criteria for NC was as follows: 1) No subjective memory complaints, 2) preserved activities of daily living and cognitive function, 3) Mini-mental state examination (MMSE) score of between 24 to 30 inclusive, 4) Clinical Dementia Rating (CDR) score of 0, and 5) education-adjusted score on delayed recall of one paragraph from Wechsler Memory Scale Logical Memory II of >=3 for 0-7 years of education, >= 5 for 8-15 years of education, and >= 9 for >=16 years of education. The criteria for MCI was as follows: 1) significant subjective memory complaints reported by the participant, clinician or informant, 2) not significantly impaired in other cognitively domains, 3) essentially preserved activities of daily living and does not meet criteria for diagnosis of dementia, 4) MMSE score of between 24 to 30 inclusive, 5) CDR score of 0.5, and 6) education-adjusted score on delayed recall of one paragraph from Wechsler Memory Scale Logical Memory II of 3-6 for 0-7 years of education, 5-9 for 8-15 years of education, and 9-11 for >=16 years of education \cite{petersen2010alzheimer}.

For the Asian disease cohort, all participants completed a locally validated neuropsychological test battery, which assessed seven domains: executive function, attention, language, visuomotor speed, verbal memory, and visual memory. Impairment in a particular domain was defined as failing at least half of the individual tests in a domain, and failure in an individual test was determined using education-adjusted cut-offs of 1.5 standard deviations below established normal means. NC was defined as having no impairment in all cognitive domains on the neuropsychological test battery, while MCI was defined as having an impairment in at least one cognitive domain of the neuropsychological test battery. Detailed descriptions of the neuropsychological assessments and diagnostic criteria are described in previous work which will be added upon acceptance.

\subsection{Amyloid +/-}

Participants from the ADNI cohort were also classified as amyloid positive or amyloid negative, using a threshold of global [18F]-Florbetapir amyloid PET SUVR >= 1.11 to define amyloid positivity \cite{petersen2010alzheimer}.

\section{Additional Implementation Details}
\label{B}
\textbf{Optimization for pre-training.} The default settings are detailed in Table \ref{tab:Pre-training setting.}. We initialized all transformer blocks using the Xavier uniform method, as described in \citep{He2022MAE}. The pre-training process utilized four A100 GPUs, each with 40GB of memory.

\textbf{Optimization for downstream tasks.} The default settings for end-to-end fine-tuning and linear probing are detailed in Table \ref{tab:End-to-end fine-tuning setting.}. For fine-tuning, following \citep{He2022MAE}, we applied layer-wise \textit{lr} decay \citep{Clark2020ELECTRA}. For linear probing, we incorporated an additional BatchNorm layer \cite{Batchnorm2015ICML} before the linear head, as per \cite{He2022MAE}.

\textbf{Masking.} The range ratios for obtaining the observation block and three target blocks introduced in Section \ref{3.2} are presented in Tables \ref{tab:Mask range of different regions.}. $\eta^a_b$ denotes the mask range along the $b$ dimension for the $a$ block. We set $K=1$ which is the number of randomly sampled blocks in three target regions.

\begin{table}[t]
  \caption{Pre-training settings. GAS: Gradient accumulation steps; BS: Batch size}
  \label{tab:Pre-training setting.}
  \renewcommand{\arraystretch}{1}
  \centering
\begin{tabular}{l|l}
\toprule
config                 & value                                   \\ \hline
optimizer              & AdamW \citep{loshchilov2018decoupled}                                   \\
optimizer momentum     & $\beta_1,\beta_2=0.9,0.999$     \\
learning rate schedule & warmup cosine schedule \citep{assran2023self}                   \\
start learning rate    & $5\times 10^{-5}$                      \\
learning rate          & $1\times 10^{-3}$                               \\
final learning rate    & $1\times 10^{-6}$                               \\
weight decay schedule  & cosine weight decay schedule \citep{assran2023self}   \\
weight decay           & 0.04                                    \\
final weight decay     & 0.4                                     \\
EMA momentum schedule     & linear \citep{assran2023self}                                     \\
EMA start momentum     & 0.996                                     \\
EMA final momentum     & 1                                     \\
total batch size       & 4 GPUs $\times$ 8 GAS $\times$ 16 BS                                 \\
warmup epochs          & 40    \\
patch size $p$          & 16 \\
dimension of gradient vector $m$       & 30 \\
training epochs  & 300 \\
\bottomrule
\end{tabular}
\end{table}

\begin{table}[t]
  \caption{Settings of end-to-end fine-tuning and linear probe.}
  \label{tab:End-to-end fine-tuning setting.}
  \renewcommand{\arraystretch}{1}
  \centering
\begin{tabular}{l|ll}
\toprule
config                 & value of FT  & value of LP  \\ \hline
optimizer              & AdamW        & LARS \citep{ginsburg2018large}         \\
optimizer momentum     & (0.9, 0.999) & 0.9          \\
learning rate schedule \ \ \ \ \ \ \ \ \ \ \ \ \ \ \ \ \ \ \ \  & cosine decay \  \citep{He2022MAE} & cosine decay \ \  \\
base learning rate     & 0.001         & 0.01         \\
weight decay           & 0.05         & N.A.             \\
layer-wise lr decay    & 0.75         & N.A.             \\
batch size             & 16           & 64           \\
warmup epochs          & 0            & 0            \\
training epochs        & 50           & 50          \\ \bottomrule
\end{tabular}
\end{table}

\begin{table}[t]
  \caption{Hyperparameters for spatiotemporal masking.}
  \label{tab:Mask range of different regions.}
  \renewcommand{\arraystretch}{1.3}
  \centering
\begin{tabular}{l|l}
\toprule
Region                 & Mask ratio                                   \\ \hline
observation block \ \ \ \ \ \ \ \ \ \ \ \ \ \ \ \ \ \ \ \ \ \ \ \ \ \ \ \ \ \ \           & \{$\eta^{\text{o}}_R$,\ $\eta^{\text{o}}_T$\}=\{(0.84,\ 1.0),\ (0.84,\ 1.0)\} \ \                               \\
target $\alpha$     & \{$\eta^{\alpha}_R$,\ $\eta^{\alpha}_T$\}=\{(0.45,\ 0.6),\ (0.2,\ 0.6)\}    \\
target $\beta$ & \{$\eta^{\beta}_R$,\ $\eta^{\beta}_T$\}=\{(0.15,\ 0.3),\ (0.0,\ 0.4)\}                   \\
target $\gamma$    & \{$\eta^{\gamma}_R$,\ $\eta^{\gamma}_T$\}=\{(0.15,\ 0.3),\ (0.0,\ 0.4)\}                     \\ \bottomrule                                 
\end{tabular}
\label{tab6}
\end{table}

\section{Additional Results}
\label{D}

\subsection{Results on additional baselines and datasets}
\label{D.1}
We incorporated more baseline results for downstream tasks on external datasets in Tables \ref{tab:Results of additional baselines on HCP-aging.}-\ref{tab:Results of additional baselines on ADNI.}, including commonly used SVM/SVR \cite{Schulz2020} and recent self-supervised learning methods. It is observed that Brain-JEPA outperforms these models on most tasks. We note that for the compared baselines, BrainMass \cite{BrainMass} is a concurrent work. Additionally, CSM \cite{CSM} and SwiFT \cite{Swift} are not time series models; CSM utilizes text-like representations, while SwiFT operates on raw fMRI data. 

\begin{table}[!h]\
\centering
\caption{Results of additional baselines on HCP-aging.
}
\label{tab:Results of additional baselines on HCP-aging.}
\renewcommand{\arraystretch}{1.5}
\resizebox{0.59\textwidth}{!}{
\begin{tabular}{lcccc}
\hline
\multirow{2}{*}{\large Methods} & \multicolumn{2}{c}{\large Age}                       & \multicolumn{2}{c}{\large Sex}                       \\ \cmidrule(lr){2-3} \cmidrule(lr){4-5}
                         & \large MSE $\downarrow$                   & \large $\rho$ $\uparrow$  & \large ACC (\%) $\uparrow$             & \large F1 (\%) $\uparrow$              \\ \hline
\large SVM/SVR                  & 0.586 (.019)          & 0.699 (.022)          & 76.67 (1.88)          & 80.82 (1.15)          \\
\large BrainMass                & 0.396 (.002)          & 0.831 (.014)          & 74.09 (3.87)          & 75.78 (3.37)          \\
\large CSM                      & 0.409 (.012)          & 0.733 (.023)          & 74.85 (1.11)          & 76.23 (0.37)          \\
\large SwiFT                    & 0.341 (.007)          & 0.755 (.063)          & 73.48 (2.20)          & 74.65 (2.32)          \\
\rowcolor{lightgray} \large \textbf{Brain-JEPA}      & \textbf{0.298} (.017) & \textbf{0.844} (.030) & \textbf{81.52} (1.03) & \textbf{84.26} (0.82) \\ \hline
\end{tabular}}
\end{table}

\begin{table}[!h]
\centering
\caption{Results of additional baselines on ADNI.
}
\label{tab:Results of additional baselines on ADNI.}
\renewcommand{\arraystretch}{1.5}
\resizebox{0.59\textwidth}{!}{
\begin{tabular}{lcccc}
\hline
\multirow{2}{*}{\large Methods} & \multicolumn{2}{c}{\large NC/MCI}                    & \multicolumn{2}{c}{\large Amyloid $a\beta+$ve/$-$ve}                    \\ \cmidrule(lr){2-3} \cmidrule(lr){4-5}
                         & \large ACC (\%) $\uparrow$             & \large F1 (\%) $\uparrow$              & \large ACC (\%) $\uparrow$             & \large F1 (\%) $\uparrow$              \\ \hline
\large SVM/SVR                  & 64.21 (5.16)          & 73.06 (4.71)          & 62.00 (4.00)          & 63.84 (5.44)          \\
\large BrainMass                & 74.21 (5.10)          & 81.36 (3.56)          & 68.00 (7.48)          & 69.29 (8.96)          \\
\large CSM                      & 68.42 (4.99)          & 76.74 (4.54)          & 63.00 (9.80)          & 65.89 (9.79)          \\
\large SwiFT                    & 73.16 (5.31)          & 80.46 (4.16)          & 65.00 (6.32)          & 67.79 (6.38)          \\
\rowcolor{lightgray} \large \textbf{Brain-JEPA}      & \textbf{76.84} (1.05) & \textbf{86.32} (0.54) & \textbf{71.00} (4.90) & \textbf{75.97} (3.93) \\ \hline
\end{tabular}}
\end{table}

To further demonstrate the diversity of our downstream applications, we conducted additional experiments using two aging-related public datasets: \href{https://sites.wustl.edu/oasisbrains/}{OASIS-3} and \href{https://cam-can.mrc-cbu.cam.ac.uk/dataset/}{CamCAN}, for AD conversion prediction in MCI participants and depression diagnosis, respectively. The results are shown in Table \ref{tab:Additional tasks of AD conversion prediction and depression classification on OASIS-3 and CamCAN datasets.}. By applying Brain-JEPA to five downstream datasets across eight distinct tasks totally, we have demonstrated its versatility in a wider range of applications compared to the existing models. Specifically, Brain-JEPA excels in demographic prediction, trait prediction, and disease diagnosis and prognosis. This stands in contrast to experiments done in BrainLM, which is limited to demographic and clinical score prediction, and BrainMass, which focuses solely on disease diagnosis and prognosis.

\begin{table}[!h]
\centering
\caption{AD conversion prediction and depression diagnosis on OASIS-3 and CamCAN datasets.
}
\label{tab:Additional tasks of AD conversion prediction and depression classification on OASIS-3 and CamCAN datasets.}
\renewcommand{\arraystretch}{1.5}
\resizebox{0.59\textwidth}{!}{
\begin{tabular}{lcccc}
\hline
\multicolumn{1}{c}{\multirow{3}{*}{\large Methods}} & \multicolumn{2}{c}{\large OASIS-3}                   & \multicolumn{2}{c}{\large CamCAN}                    \\ \cmidrule(lr){2-3} \cmidrule(lr){4-5} 
\multicolumn{1}{l}{}                         & \multicolumn{2}{c}{\large AD Conversion}            & \multicolumn{2}{c}{\large Depression}              \\ \cmidrule(lr){2-3} \cmidrule(lr){4-5}
\multicolumn{1}{l}{}                         & \large ACC (\%) $\uparrow$              & \large F1 (\%) $\uparrow$               & \large ACC (\%) $\uparrow$              & \large F1 (\%) $\uparrow$               \\ \hline
\large SVM/SVR                                      & 56.00 (2.81)          & 52.05 (1.66)          & 63.64 (3.07)          & 56.79 (2.32)          \\
\large BrainNetCNN                                  & 62.00 (2.45)          & 59.53 (0.58)          & 62.73 (4.45)          & 56.85 (4.47)          \\
\large BrainGNN                                     & 59.00 (2.00)          & 56.53 (4.34)          & 63.64 (4.98)          & 56.68 (3.26)          \\
\large BNT                                          & 68.00 (8.72)          & 64.73 (11.29)         & 65.45 (4.64)          & 55.32 (8.67)          \\
\large BrainLM                                      & 65.00 (7.75)          & 62.67 (9.04)          & 70.00 (6.17)          & 64.18 (3.82)          \\
\large BrainMass                                    & 67.00 (6.00)          & 66.53 (6.95)          & 70.91 (2.23)          & 63.56 (2.93)          \\
\large CSM                                          & 61.00 (4.90)          & 61.97 (5.49)          & 64.55 (4.45)          & 56.08 (6.23)          \\
\large SwiFT                                        & 65.00 (6.32)          & 66.80 (4.12)          & 69.09 (6.68)          & 61.78 (9.26)          \\
\rowcolor{lightgray} \large \textbf{Brain-JEPA}                          & \textbf{69.00} (7.35) & \textbf{67.32} (7.92) & \textbf{72.73} (2.87) & \textbf{67.45} (1.57) \\ \hline
\end{tabular}}
\end{table}

\subsection{Scaling properties with respect to dataset size.}
\label{D.3}
We compared the performance of Brain-JEPA trained with varying portions of the UKB pretraining dataset: 25\%, 50\%, 75\%, and 100\%. As shown in Table \ref{portion}, the performance improves as the dataset size increases, highlighting the scalability of Brain-JEPA in relation to the pretraining dataset size.

\begin{table}[!h]
\centering
\caption{Ablation on different dataset size for pretraining.
}
\renewcommand{\arraystretch}{1.5}
\resizebox{0.45\textwidth}{!}{
\begin{tabular}{lccc}
\hline
\multirow{3}{*}{\large Methods} & \multicolumn{2}{c}{\large HCP-Aging}      & \large ADNI         \\ \cline{2-4} 
                         & \large Age                 & \large Sex          & \large NC/MCI       \\
                         &\large  $\rho$ $\uparrow$ &\large  ACC (\%) $\uparrow$    &\large  ACC (\%) $\uparrow$    \\ \hline
\large 25\%                     & 0.659 (.043)        & 68.03 (1.21) & 67.89 (9.18) \\
\large 50\%                     & 0.768 (.012)        & 74.24 (1.36) & 71.05 (3.86) \\
\large 75\%                     & 0.813 (.015)        & 77.42 (2.00) & 74.74 (4.88) \\
\large 100\%                    & \textbf{0.844} (.030)        & \textbf{81.52} (1.03) & \textbf{76.84} (1.05) \\ \hline
\end{tabular}}
\label{portion}
\end{table}

\subsection{Additional ablations}
\label{D.2}
\textbf{Architectures/Frameworks}. To thoroughly compare the performance between JEPA with anatomical locations (AL) and BrainLM (MAE-based), we extended our comparison to include all the tasks except for the three in the main content, as well as two newly added datasets, OASIS-3 and CamCAN. The results shown in Table \ref{tab:Ablation on position embedding.}, demonstrates that JEPA with AL outperforms BrainLM in seven out of eleven tasks, demonstrating the superiority of prediction in latent space. For the tasks where BrainLM performs better, it is likely that JEPA requires gradient positioning for precise ROI placement to achieve optimal performance. In future work, we will further investigate the possible interactions between the self-supervised learning framework and brain gradient positioning.
\begin{table}[!h]
\centering
\caption{Ablation on position embedding.
}
\label{tab:Ablation on position embedding.}
\renewcommand{\arraystretch}{1.5}
\resizebox{0.59\textwidth}{!}{
\begin{tabular}{lcccc} \\ \hline
\multicolumn{1}{l}{\multirow{3}{*}{\large Methods}}      & \multicolumn{2}{c}{\large UKB}                   & \multicolumn{2}{c}{\large HCP-Aging}                    \\ \cmidrule(lr){2-3} \cmidrule(lr){4-5}
                 &\large Age                    &\large  Sex                   &\large Neurotism             &\large Flanker               \\ \cmidrule(lr){2-3} \cmidrule(lr){4-5}
                 & \large $\rho$ $\uparrow$                       &\large  ACC (\%) $\uparrow$            & \large $\rho$ $\uparrow$                      & \large $\rho$ $\uparrow$                      \\ \hline
\large BrainLM          & 0.632 (0.020)          & \textbf{86.47} (0.74) & 0.231 (.012)          & 0.318 (.048)          \\
\large Brain-JEPA w AL  & \textbf{0.686} (0.013) & 84.11 (0.50)          & \textbf{0.267} (.003) & \textbf{0.374} (.022) \\ \hline
\multicolumn{1}{l}{\multirow{3}{*}{\large Methods}}        & \large ADNI          & \large MACC        & \large  OASIS-3      & \large CamCAN       \\ \cmidrule(lr){2-5}
                 & \large Amy+/-                 & \large NC/MCI                &\large  AD Conversion         &\large  Depression            \\ \cmidrule(lr){2-5}
                 & \large ACC (\%) $\uparrow$              &\large  ACC (\%) $\uparrow$             & \large ACC (\%) $\uparrow$             &\large  ACC (\%) $\uparrow$             \\ \hline
\large BrainLM & \textbf{67.00} (7.48)  & 61.65 (3.35)          & 65.00 (7.75)          & 70.00 (6.17)          \\
\large Brain-JEPA \textit{w} AL  & 65.00 (6.32)           & \textbf{64.33} (1.80)& \textbf{67.00} (4.00) & \textbf{71.82} (6.03) \\ \hline
\end{tabular}}
\end{table}

We further compared Brain-JEPA without JEPA architecture (\textit{i.e.}, BrainLM with contributions) to Brain-JEPA. As shown in Table \ref{tab:Comparisons of different frameworks.}, Brain-JEPA (JEPA framework) outperforms BrainLM (MAE framework) with contributions consistently, demonstrating the superiority of JEPA framework.

\begin{table}[!h]
\centering
\caption{Comparisons of different frameworks.
}
\label{tab:Comparisons of different frameworks.}
\renewcommand{\arraystretch}{1.5}
\resizebox{0.59\textwidth}{!}{
\begin{tabular}{lccc}
\hline
\multirow{3}{*}{\large Methods} & \multicolumn{2}{c}{\large HCP-Aging}                 &\large  ADNI                  \\ \cline{2-4} 
                         &\large  Age                   &\large  Sex                   &\large  Amy+/-                \\ \cline{2-4} 
                         & \large $\rho$ $\uparrow$                       &\large  ACC (\%) $\uparrow$              & \large ACC (\%) $\uparrow$              \\ \hline
\large BrainLM                  & 0.832 (.028)          & 74.39 (1.55)          & 67.00 (7.48)          \\
\large BrainLM w contributions  & 0.838 (.014)         & 76.36 (2.58)          & 70.00 (11.40)         \\
\large JEPA w contributions     & \textbf{0.844} (.030) & \textbf{81.52} (1.03) & \textbf{71.00} (4.90) \\ \hline
\end{tabular}}
\end{table}

\textbf{The number of gradient components.} We compared the model performance between 3-dimensional (3-dim) and 30-dim brain gradient positioning, shown in Table \ref{dim}. The 30-dim model consistently outperformed the 3-dim model by a large margin. This indicates that higher-dimensional brain gradients may encapsulate finer-grained information on brain network organization, which benefits the learning of brain dynamics.

\begin{table}[!h]
\centering
\caption{Comparison of different number of gradient components.
}
\label{tab:Comparison of different numbers of gradient components.}
\renewcommand{\arraystretch}{1.5}
\resizebox{0.59\textwidth}{!}{
\begin{tabular}{lclc}
\hline
\multirow{3}{*}{\large Methods} & \multicolumn{2}{c}{\large HCP-Aging}                                            &\large  ADNI                                      \\ \cline{2-4} 
                         &\large  Age                                       & \multicolumn{1}{c}{\large Sex}      &\large  Amy+/-                                    \\ \cline{2-4} 
                         &  \large $\rho$ $\uparrow$                                         &\multicolumn{1}{c}{\large  ACC (\%)} $\uparrow$ &\large  ACC (\%) $\uparrow$                                  \\ \hline
\large 3-dim brain gradient                 & \multicolumn{1}{l}{0.819} (.003)         & 76.96 (1.77)                 & \multicolumn{1}{l}{67.00 (6.00)}         \\
\large 30-dim brain gradient  \ \ \ \ \ \ \              & \multicolumn{1}{l}{\textbf{0.844} (.030)} & \textbf{81.52} (1.03)        & \multicolumn{1}{l}{\textbf{71.00} (4.90)} \\ \hline
\end{tabular}}
\label{dim}
\end{table}

\section{Broader Impact}
\label{C}

The introduction of Brain-JEPA marks a significant advancement in the interdisciplinary field of AI and neuroscience, particularly in the brain activity analysis. An assessment of the broader impact of this model has across various dimensions:

\subsection{Neuroscience and medical advancements}

Brain-JEPA's capabilities in demographic prediction, disease diagnosis, and prognosis could revolutionize how neurological disorders are diagnosed and treated. This may lead to earlier detection and more personalized therapeutic interventions, potentially improving outcomes for patients with conditions like AD, schizophrenia, or autism spectrum disorders. Furthermore, the model's innovative techniques, including Brain Gradient Positioning and Spatiotemporal Masking, offer new ways to understand the brain's functional organization. This could lead to breakthroughs in identifying how various cognitive processes are mapped in the brain, aiding in both basic science and clinical applications. On the other hand, by effectively predicting various traits, Brain-JEPA can aid in the study of the genetic and environmental influences on behavior and cognitive functions. This can enhance our understanding of the neural underpinnings of psychological traits and disorders.

\subsection{Technological impact}

Brain-JEPA sets a new standard in AI's application to complex brain activity data with a novel brain functional coordinate system and masking strategy, which could spur further innovations and applications of AI across different sub-fields of neuroscience. Furthermore, the model's success in performing well across different ethnic groups indicate potential for broad applications in diverse global settings, which is crucial for building inclusive and unbiased AI systems.

\subsection{Ethical and social considerations}

Ensuring the confidentiality and integrity of patient data while using such advanced AI systems is paramount. While Brain-JEPA has shown superior performance across different tasks, continuous monitoring for potential biases is essential, especially as the model is scaled and deployed in varied clinical settings. Besides, the deployment of advanced technologies like Brain-JEPA could exacerbate existing disparities in healthcare access unless carefully managed. Ensuring that these technologies benefit all segments of the population equally is critical.


\newpage

\section*{NeurIPS Paper Checklist}

\begin{enumerate}

\item {\bf Claims}
    \item[] Question: Do the main claims made in the abstract and introduction accurately reflect the paper's contributions and scope?
    \item[] Answer: \answerYes{} 
    \item[] Justification: Refer to Section \ref{intro} for our contributions and scope.

\item {\bf Limitations}
    \item[] Question: Does the paper discuss the limitations of the work performed by the authors?
    \item[] Answer: \answerYes{} 
    \item[] Justification: Refer to Section \ref{limitation} for the discussion of limitation and future work.

\item {\bf Theory Assumptions and Proofs}
    \item[] Question: For each theoretical result, does the paper provide the full set of assumptions and a complete (and correct) proof?
    \item[] Answer: \answerNA{} 
    \item[] Justification: The paper does not include theoretical results.

    \item {\bf Experimental Result Reproducibility}
    \item[] Question: Does the paper fully disclose all the information needed to reproduce the main experimental results of the paper to the extent that it affects the main claims and/or conclusions of the paper (regardless of whether the code and data are provided or not)?
    \item[] Answer: \answerYes{} 
    \item[] Justification: We have provided detailed information about the datasets in Section \ref{dataset} and Appendix \ref{A}. The implementation details are presented in Section \ref{implementation} and Appendix \ref{B}. The model checkpoints and anonymized code are provided in supplementary material with clear instructions in the readme file.

\item {\bf Open access to data and code}
    \item[] Question: Does the paper provide open access to the data and code, with sufficient instructions to faithfully reproduce the main experimental results, as described in supplemental material?
    \item[] Answer: \answerYes{} 
    \item[] Justification: We have provided model checkpoints and code in supplementary material with a readme file. The links to the publicly available datasets are provided.

\item {\bf Experimental Setting/Details}
    \item[] Question: Does the paper specify all the training and test details (e.g., data splits, hyperparameters, how they were chosen, type of optimizer, etc.) necessary to understand the results?
    \item[] Answer: \answerYes{} 
    \item[] Justification: We have provided detailed information about the datasets in Section \ref{dataset} and Appendix \ref{A}.  We have also provided the implementation details and hyperparameters in Section \ref{implementation} and Appendix \ref{B}.

\item {\bf Experiment Statistical Significance}
    \item[] Question: Does the paper report error bars suitably and correctly defined or other appropriate information about the statistical significance of the experiments?
    \item[] Answer: \answerYes{} 
    \item[] Justification: All main results in Table \ref{tab1}, \ref{tab2}, and \ref{tab3} contain mean and standard deviation from 5 independent runs. The statistical significance of the experiments has been shown with \textbf{*} denoting significant improvement with $p<0.05$

\item {\bf Experiments Compute Resources}
    \item[] Question: For each experiment, does the paper provide sufficient information on the computer resources (type of compute workers, memory, time of execution) needed to reproduce the experiments?
    \item[] Answer: \answerYes{} 
    \item[] Justification: We provide the computer resources in Section \ref{implementation}.
    
\item {\bf Code Of Ethics}
    \item[] Question: Does the research conducted in the paper conform, in every respect, with the NeurIPS Code of Ethics \url{https://neurips.cc/public/EthicsGuidelines}?
    \item[] Answer: \answerYes{} 
    \item[] Justification: The proposed research conforms with the Code of Ethics.

\item {\bf Broader Impacts}
    \item[] Question: Does the paper discuss both potential positive societal impacts and negative societal impacts of the work performed?
    \item[] Answer: \answerYes{} 
    \item[] Justification: Refer to Section \ref{C} for  potential positive societal impacts and negative societal impacts of Brain-JEPA.
    
\item {\bf Safeguards}
    \item[] Question: Does the paper describe safeguards that have been put in place for responsible release of data or models that have a high risk for misuse (e.g., pretrained language models, image generators, or scraped datasets)?
    \item[] Answer: \answerNA{} 
    \item[] Justification: The paper poses no such risks.

\item {\bf Licenses for existing assets}
    \item[] Question: Are the creators or original owners of assets (e.g., code, data, models), used in the paper, properly credited and are the license and terms of use explicitly mentioned and properly respected?
    \item[] Answer: \answerYes{} 
    \item[] Justification: The original papers that produced the code package or dataset are all properly cited.

\item {\bf New Assets}
    \item[] Question: Are new assets introduced in the paper well documented and is the documentation provided alongside the assets?
    \item[] Answer: \answerYes{} 
    \item[] Justification: The details of the dataset/code/model have been provided in \ref{dataset}, \ref{implementation} and Appendix \ref{B}, with the codes in supplementary materials.

\item {\bf Crowdsourcing and Research with Human Subjects}
    \item[] Question: For crowdsourcing experiments and research with human subjects, does the paper include the full text of instructions given to participants and screenshots, if applicable, as well as details about compensation (if any)? 
    \item[] Answer: \answerNA{} 
    \item[] Justification: The datasets of human subjects used in this paper are either publicly available or from previous work which will be cited upon acceptance.

\item {\bf Institutional Review Board (IRB) Approvals or Equivalent for Research with Human Subjects}
    \item[] Question: Does the paper describe potential risks incurred by study participants, whether such risks were disclosed to the subjects, and whether Institutional Review Board (IRB) approvals (or an equivalent approval/review based on the requirements of your country or institution) were obtained?
    \item[] Answer: \answerNA{} 
    \item[] Justification: The datasets of human subjects used in this paper are either publicly available or from previous work which will be cited upon acceptance.

\end{enumerate}

\end{document}